\documentstyle[aasms,12pt]{article}

\input psfig.tex
\def\etal{{\it et al.}}
\def\mxth{\mathsurround=0pt }
\def\VEV#1{\left\langle #1\right\rangle}
\def\xversim#1#2{\lower2.pt\vbox{\baselineskip0pt \lineskip-.2pt
    \ialign{$\mxth#1\hfil##\hfil$\crcr#2\crcr\sim\crcr}}}

\def\lta{\mathrel{\mathpalette\xversim <}}
\def\gta{\mathrel{\mathpalette\xversim >}}
\def\etal{{\it et al.}}
\def\ref@jnl#1{{\jnl@style#1}}
\def\nuphysb{\ref@jnl{Nuc. Phys. B}}
\begin{document}
\begin{center}
{\bf POLARIZATION OF THE MICROWAVE BACKGROUND DUE TO PRIMORDIAL
GRAVITATIONAL WAVES}
\end{center}

\vspace{.2in}
\begin{center}
Robert Crittenden, Richard L. Davis, and Paul J. Steinhardt \\
{\it Department of Physics, University of Pennsylvania,
Philadelphia, PA  19104}
\end{center}

\begin{quote}
{\bf Abstract:}
The contribution of gravitational wave (tensor metric)
and energy density (scalar metric)
fluctuations to the cosmic microwave background polarization
is computed.
We find that the tensor contribution is significant
 only at
 large angular scales (multipoles $\ell \lta 40$).  For standard
 recombination, the tensor contribution can  dominate  at
 $\ell \lta 40$; however, the effect would be difficult to detect since
 the total (scalar plus tensor) polarization is $< 1$\%.
 For models with late reionization, the total large angular scale
 polarization is
 large ($\sim 7-9$\%), but the tensor fraction is negligibly small.
 Hence, polarization may be useful for discriminating ionization
 history, but is less promising as a means for detecting tensor fluctuations.
 \end{quote}

 \noindent
{\it Subject headings}:
cosmology: cosmic microwave background --- cosmology: observations

\vspace{.3in}

The cosmic microwave background (CMB) temperature anisotropy may be
induced by a combination of energy density (\cite{bardeen83})
(scalar metric)
fluctuations and gravitational wave (\cite{starobinsky85}, \cite{abbott84})
(tensor metric) fluctuations.
Distinguishing the scalar and tensor components is important
for testing
cosmological models
(\cite{davis92}, \cite{krauss92}, \cite{salopek92},
\cite{sahni92}, \cite{liddle92}, \cite{lucchin92}, \cite{lidsey92}),
since only scalar metric fluctuations can grow into non-linear structures,
such as galaxies, while tensor fluctuations red shift and disperse upon
entering
the horizon.
One  suggested approach for determining the
tensor contribution is to compare measurements of
the CMB anisotropy on  scales ranging from
large ($\lta1^{\circ}$) to small ($\gta 1^{\circ}$) and take advantage of the
fact that unpolarized
scalar and tensor  fluctuations
vary differently with angular scale (\cite{davis92},\cite{crittenden93}).
Another suggestion has been to measure the CMB polarization
since tensor fluctuations
might  induce a polarization significantly greater than a spectrum
 of purely energy density perturbations (\cite{polnarev85}, \cite{ng93}).

 In this Letter we report on a computation of the polarization induced by a
primordial
 spectrum of energy density and gravitational wave perturbations.
By
 solving numerically the radiation transfer equations, we find
 that tensor perturbations significantly enhance the net polarization
 at large angular scales (multipoles $\ell \lta 40$).
   For example,
 in the case where there are
 equal scalar and tensor contributions to the $\Delta T/T$ (unpolarized)
 quadrupole moment,
 the tensor contribution can boost the polarization quadrupole
 by a factor of four or more
  compared to models with scalar fluctuations only.
 Nevertheless, we find the prospects for detecting
primordial gravitational radiation using polarization are not promising:
 For standard recombination
 the net polarization on large-angular$(\gta 1^{\circ})$ scales
  is less than 1\% of $\Delta T/T$, too small
 to be detected with current technology.  On small angular scales,  the net
 polarization rises, but the tensor fraction
 drops  to undetectable values.
 The fractional decrease occurs  because
 the
 tensor contribution to $\Delta T/T$
 on subdegree scales diminishes rapidly with
 increasing $\ell$ whereas the scalar anisotropy
 rises (\cite{starobinsky85}, \cite{crittenden93}).
 Changing the Hubble constant and/or introducing a
 cosmological constant does not significantly alter the conclusions.
 For nonstandard reionization histories, the net polarization  can be much
 higher ($\sim 7-9$\%).  Hence, polarization may be a useful tool for
testing ionization history.  However, now the tensor fraction does not
dominate on any angular scale, so that polarization is still not  useful
discriminating tensor fluctuations.

Our results apply to the general class of flat, Friedmann-Robertson-Walker
universes with a gaussian distribution of adiabatic initial fluctuations,
assuming
power-law spectra and cold dark matter.  An important
subset of these models have strong theoretical motivation
from inflationary cosmology.
Inflation predicts a nearly scale-invariant
(spectral index $n \approx 1$) spectrum of both scalar and tensor
fluctuations.
 (The scalar fluctuations are
created by quantum fluctuations of the inflaton field which drives
the transition from the false to true vacuum phase (\cite{bardeen83}).
Quantum fluctuations in the inflaton on microscopic scales are
stretched by inflation into  a spectrum of
energy density perturbations spanning cosmological wavelengths.
Similarly, tensor fluctuations  are produced by quantum fluctuations of
the graviton field which are   inflated to cosmological
wavelengths (\cite{starobinsky85},  \cite{abbott84}).
The relative contributions of scalar and tensor
fluctuations to the CMB anisotropy depends upon the details of the
inflationary potential.  Likewise, the  spectral tilt, $1-n$, is
model dependent. However,  there is a nearly
model-independent relation
between the two:
\begin{equation}
C^{(T)}_2/ C^{(S)}_2 \approx 7(1-n) \ ,
\end{equation}
where $C^{(T)}_2$ and $C^{(S)}_2$ are the $\ell =2$ (quadrupole)
moments of the power spectrum
(\cite{davis92},
\cite{sahni92}, \cite{liddle92}, \cite{lucchin92}, \cite{lidsey92}).
(The only known exceptions to Eq.~(1) require artificial, exponential
fine-tuning of inflationary potential parameters.)
 A key test for inflation is  to measure independently
 the ratio of
 tensor to scalar contributions to the CMB anisotropy
 and the tilt, $1-n$, and determine if they satisfy the relation
 above.

As indicated in the preceding, there is
 strong motivation for investigating
whether polarization
 measurements can be useful in extracting the tensor component of
 the CMB anisotropy.  In order
to compute the scalar- and tensor-induced polarization we evolve
the photon distribution function, ${\bf f} ({\bf x}, {\bf q}, t) $,
for photons at position ${\bf x}$ at
time $t$ with momentum ${\bf q}$, using first-order perturbation
theory of the general relativistic Boltzmann equation for radiative
transfer (\cite{bond84}, \cite{bond87})
with a Thomson scattering source term, in the synchronous gauge.
Photon
polarization is included by making ${\bf f}$ a 4-dimensional vector
with components related to the Stokes parameters ($f_s$ with
$s=t,p,u,v$ correspond to the usual $I,Q,U,V$ Stokes notation), where
we use
Chandrasekhar's treatment of the scattering source term for
Rayleigh (and thus Thomson) scattering in a plane parallel
atmosphere (\cite{chandrasekhar60}).
In the scalar case, only $f_t$ and
the `polarization' $f_p$ are needed, so two transfer equations are
required (\cite{bond84}). In the tensor case $f_u$ also does not vanish, but
it is related to $f_p$, so again only two perturbed transfer equations
turn out to be required, the second of which describes the
polarization. The full radiation transfer equations (coupled equations
for the unpolarized and polarized anisotropy) are given
in \cite{bond84} for the scalar case and \cite{crittenden93}
for the tensor case.
By numerically solving these equations, we have extracted  the
tensor and scalar multipole moments, $C_{\ell}^{(S)}$, $C_{\ell}^{(S)P}$, and
$C_{\ell}^{(T)}$, $C_{\ell}^{(T)P}$,
 for both unpolarized and polarized measurements.
The sum
is the total multipole moment, $C_{\ell}$, which describes
the  temperature autocorrelation function:
\begin{equation}
C(\alpha) = \VEV{\frac{\Delta T}{T}({\bf q}) \frac{\Delta T}{T}({\bf q}')} =
\sum_{\ell} (2 \ell +1) C_{\ell} P_{\ell} ({\rm cos}\;\alpha),
\end{equation}
where ${\bf q} \cdot {\bf q}' = \; \cos{\alpha}$.

Fig.~1 shows  $C_{\ell}^{P}/C_{\ell}$, the ratio of
polarized to unpolarized moments, vs. $\ell$.
Since moment $\ell$ is dominated by
fluctuations on angular scales $\approx \pi/\ell$,  we can use
the square root of this  ratio
 to estimate the
 polarization-to-anisotropy ratio on angular
 scale $\approx \pi/\ell$.
  The figure corresponds  to the inflationary
prediction, Eq.~1,
with  $n=0.85$, for which  tensor and scalar contributions
to the CMB quadrupole are equal.
 The   Hubble constant is
$H_0 = 100 h$~ km/sec-Mpc with  $h=0.5$ and
the baryon density is fixed by
nucleosynthesis (\cite{walker91}), $\Omega_B=0.125h^{-2}$.
As stated above, we assume that
$\Omega_{total} = 1$, where the missing mass is cold dark matter.
The results  given here are relatively insensitive
 to
variations in $h$
or the  cosmological constant  $ \Lambda$, or a possible hot
component.  Quantitatively, the polarization changes by $\ll 1$\%
for variations $0.5 \le h \le 1.0$ and $0.0 \le \Omega_{\Lambda}\le 0.8$.

 In Fig.~1,
 the tensor contribution to the polarization  significantly dominates the
 scalar contribution
 for  $\ell \lta 40$ corresponding to length
 scales that were outside the horizon at decoupling.
 However, over this range,
 the  net polarization-to-anisotropy ratio is
 less than  1\%.  Since  the CMB anisotropy is
 $\Delta T/T \sim O(10^{-5})$, a polarization experiment with
 sensitivity  $ < O(10^{-7})$ is needed to detect
 the tensor-induced polarization over this range.  The
 requisite sensitivity
 is at the limit of  the most optimistic estimates for
   near-future detectors (\cite{smoot}).   At subdegree scales ($l>100$), the
   polarization-to-anisotropy ratio increases,
   but, for such $\ell$'s,
   the tensor contribution to the polarization is negligible.
  Polarization and anisotropy at subdegree scales is dominated by
  wavelengths inside the horizon at decoupling.  For these wavelengths
  the gravitational wave contribution has red-shifted and dispersed
   whereas the scalar contribution has grown,
   as shown in Fig.~2.
   Consequently, the
   tensor fraction of the polarization (and anisotropy) is negligible.
   Polarization does not appear to be a  promising approach for detecting the
   gravitational wave contribution predicted by inflation and
   standard recombination.

The total polarization is enhanced if the universe is reionized at
some time after decoupling  (\cite{polnarev85}).
Fig.~3 shows the results for the case of no recombination.
  All other
 spectral and cosmological parameters are the same as in Fig.~1.
 The polarization from both scalar and tensor components (and the
 total polarization) have increased  on large angular scales to $\sim 8$\%
  (see also \cite{bond84} and \cite{nasel87}),
 levels
that may be detectable in the near future.
It appears that polarization may be useful for discriminating the
ionization history.
 However,  we also note that the tensor fraction of the polarization
 for  no recombination
 is  smaller than the scalar even for small $\ell$.
 We have also considered models with recombination followed by reionization
 at some red shift $z_r$.  Changing $z_r$ can alter the total polarization
 (\cite{stark81}).
 for $h=0.5$ and $\Omega_B=0.5$, the polarization rises to
 $\sim 9$\% for $z_r\sim 200$, corresponding to  two or three optical
 depths between reionization and the present.
 Despite modest differences in total polarization,
 the situation for no
 recombination and the non-standard reionization histories is the same in
 terms of the tensor mode: distinguishing the tensor contribution
 requires sensitivity to  better than 1

Our calculations apply to a larger class of models than those predicted
by inflation.  If we relax the inflationary predictions relating
the gravitational wave background to the spectral index, and were
to assume $a\ priori$ a primordial stochastic background of gravitational
radiation,
polarization is still not a practical
 method for extracting the tensor contribution to CMB anisotropy.
 Fig.~4 illustrates the extreme limits of pure tensor and pure
 scalar fluctuations. (In the tensor curve,
 the polarization does not diminish  at
 large $\ell$, as it does in Fig.~1,
 because the scalar anisotropy, which normally
 dominates at large $\ell$ has, been set to zero.)
The polarization-to-anisotropy  ratio
 is only weakly $n$-dependent.  For tensor fluctuations,
 the only discernible $n$-dependence  is
 on large angular scales;  for scalar fluctuations,
 there is no discernible difference on large or small scales for the range
 of $n$ illustrated in the figure.  The weak $n$-dependence makes it easy
 estimate the polarization for a wide variety of models.
 As in earlier examples, the  tensor polarization
 dominates the scalar at large angular scales, but the magnitude
 is too small to be detected in either case.
 Now we
 observe that the pure scalar and pure tensor polarizations are quite
 similar at small angular scales, too, where the polarization might be
 large enough to be detected.
 Only around $1^{\circ}$, for $\ell \approx 100$, does there appear to be
 a narrow window in which the predictions from these {\it extreme} models
 approach  measurable difference.  This difference disappears if a
 scalar contribution  greater than $\approx 10$\% is added to the pure
tensor fluctuations.

  Actual experiments
 measure a wide range of angular scales and measure many multipoles
 at the same time.  The experimental sensitivity  to different $\ell$'s
 can be expressed in
 terms of a filter function, $W_\ell$,  where:
 \begin{equation}
C(\alpha) = \sum_{\ell} (2 \ell +1) C_{\ell}
W_{\ell} P_{\ell} ({\rm cos}\;\alpha).
\end{equation}
For the purposes of illustration, we will compute the polarization
measured for the models in Figs.~1 and~3
for  a hypothetical single beam anisotropy experiment.  We shall suppose
that the experiment measures  with equal sensitivity
($W_{\ell}=1) $ all multipoles ranging from
$\ell \approx 5$ (lower $\ell$'s excluded by limited sky coverage)
to some $\ell_{max}= (50, \; 100,\; 300,\;1000)$ (fixed by the beam
size).  With these windows, we find the rms total polarization-to-anisotropy
ratio $[C_p(\alpha=0)/C(\alpha=0)]^{1/2}$ for the standard
recombination model with $n=0.85$ and
$C^{(T)}_2/ C^{(S)}_2=1$ in Fig.~1
is $(.5\%,\;1.0\%,\; 1.3\%,\; 5.4\%)$, to be compared with
$(.4\%,\;1.0\%,\; 1.4\%,\; 6.0\%)$ for a
model with $n=0.85$ but purely scalar fluctuations.
 For the non-standard
recombination model
in Fig.~3, the result for
$n=0.85$ and
$C^{(T)}_2/ C^{(S)}_2=1$
is $(7.4\%, \; 7.4\%, \; 7.4\%, 7.4\%)$,
to be compared with $(7.9\%, \; 7.8\%, \; 7.8\%, \;7.8\% )$ for a model
with same ionization history and $n=0.85$ but purely scalar fluctuations.
Distinguishing a tensor contribution in either example requires
sensitivity at the level $\ll 1$\%.  In addition,
 we find that the polarization
 can change by more than $.1$\%   if
$h$, $\Omega_B$, or $z_r$  are varied within present observational
limits. Hence,
discriminating a tensor contribution is further complicated by
uncertainties in these  cosmological parameters.

In sum, polarization may be a useful discriminant for
determining the ionization history of the universe.  However, the polarization
enhancement due to gravitational waves is too small to be resolved unless
there is more than an order of magnitude improvement in experimental
sensitivity.
Our conclusions
apply to a large class of cosmological models and cosmological parameters.
If inflation is correct, or if a primordial background
of gravitational radiation exists for some other reason, then it
appears that the most promising
method for separating the
scalar and tensor contributions to the CMB anisotropy is to combine
small- and large-angular scale
unpolarized measurements, as described by
\cite{davis92} and \cite{crittenden93}.

We thank J.R.  Bond,
G. Efstathiou,  and G. F. Smoot for useful contributions to this work.
This research was supported by the DOE at Penn (DOE-EY-76-C-02-3071).

\newpage
\begin{figure}
\caption{
Ratio of polarization multipole  to (unpolarized)
anisotropy multipole vs. moment ($\ell$)
 predicted by an inflationary model with $n=0.85$, $h=0.5$, cold
 dark matter, and standard
 recombination.  (Inflation predicts equal scalar and tensor
 contributions to the unpolarized quadrupole.)  The ratio above
 represents, roughly, the square of the polarization-to-anisotropy
 ratio on angular scale $\approx \pi/\ell$.}
\end{figure}
\begin{figure}
\caption{
The total, tensor and scalar contributions to the unpolarized
anisotropy for the model in Fig. 1.  Although the scalar and
tensor contributions to the quadrupole ($\ell=2$) moment are
equal, the tensor contribution drops to insignificant values at
subdegree ($\ell > 100$) scales, which explains why the
tensor contribution to the polarization is small at subdegree
scales in Fig. 1.}
\end{figure}


\begin{figure}
\caption{
Same as Fig.~1, except with no recombination (i.e., extended
ionization).  The net polarization is greatly enhanced on large
angular scales, but the tensor contribution is subdominant for
all $\ell$.}
\end{figure}
\begin{figure}
\caption{
Ratio of polarization moments to anisotropy moments for pure tensor
and pure
scalar spectra, with $h=0.5$, cold dark matter, and standard recombination.
 There is no discernible variation with $n$ for the pure scalar case
  and only differences at small $\ell$ for the tensor case.
  }
  \end{figure}


\clearpage

\begin{thebibliography}
\bibitem[Abbott and Wise 1984]{abbott84}\reference  Abbott, L. F. and
 Wise, M. 1984,   Nuc. Phys. B, 244, 541
\bibitem[Bardeen \etal\ 1983]{bardeen83}\reference
Bardeen J. M., Steinhardt P. J., and Turner M. S.  1983,
\prd, 28, 679
\bibitem[Bond and Efstathiou 1984]{bond84}\reference
Bond, J. R. and  Efstathiou, G. 1984,
 \apj  285, L45
 \bibitem[Bond  and Efstathiou 1987]{bond87}\reference
 Bond, J. R. and  Efstathiou, G. 1987,
 \mnras, 226, 655
 \bibitem[Chandrasekhar 1960]{chandrasekhar60}\reference
  Chandrasekhar, S. 1960, {\bf Radiation Transfer}
   (Dover: New York), 1
\bibitem[Crittenden  \etal\ 1993]{crittenden93}\reference
Crittenden, R., Bond, J. R., Davis, R. L., Efstathiou, G., and
Steinhardt, P. J. 1993, Penn preprint
\bibitem[Davis \etal\ 1992]{davis92}\reference
 Davis, R. L., Hodges, H. M.,   Smoot, G. F.,  Steinhardt, P. J.,   and
  Turner, M. S. 1992,   \prl,  69, 1856
\bibitem[Krauss and White 1992]{krauss92}\reference
Krauss, L. M., and White, M. 1992,
Phys. Rev. Lett. 69 (1992) 869-872.
\bibitem[Liddle and Lyth 1992]{liddle92}\reference
    Liddle,  A.  and  Lyth, D.  1992, Phys. Lett. B, 291, 391
\bibitem[Lidsey and Coles 1992]{lidsey92}\reference
    Lidsey, J. E.  and  Coles, P. 1992, \mnras, 358, 57
 \bibitem[Lucchin \etal\ 1992]{lucchin92}\reference
 Lucchin, F.,    Matarrese, S.,
  and Mollerach,  S. 1992, \apjl, 401, 49
\bibitem[Ng and Ng 1993]{ng93}\reference
  Ng, K. L. and Ng, K. W. 1993, Institute of Physics,
  Academica Sinica Preprint IP-ASTP-08-93
\bibitem[Polnarev 1985]{polnarev85}\reference
    Polnarev, A. G. 1985,  \sovast,  29, 607
\bibitem[Sahni and Souradeep 1992]{sahni92}\reference
      Sahni, V. and  Souradeep,  T. 1992, Mod. Phys. Lett., A7, 3541
\bibitem[Salopek 1992]{salopek92}\reference
     Salopek, D. 1992, \prl 69, 3602
\bibitem[Smoot 1993]{smoot} Smoot, G. F., private communication
\bibitem[Starobinsky 1985]{starobinsky85}\reference
 Starobinsky, A. I. 1985, Sov. Astron. Lett., 11, 133
\bibitem[Walker \etal 1991]{walker91}  Walker,  T. P.,
 Steigman, G.,
 Schramm, D. N.,  Olive, K. A., and  Kang, H. S. 1991,
\apj  376,  51
\bibitem[Nasel'skii and Polnarev 1987]{nasel87}\reference
Nasel'skii, P. D. and Polnarev, A. G. 1987, {\it Astrofizika} 16, 543
\bibitem[Stark 1981]{stark81}\reference Stark, R.F. 1981, \mnras 195, 127
\end{thebibliography}
\end{document}